# Surface optomechanics: Calculating optically excited acoustical whispering gallery modes in microspheres

John Zehnpfennig[1,2],* Gaurav Bahl[2], Matthew Tomes[2], and Tal Carmon[2]

[1.]*United States Military Academy, Electrical Engineering and Computer Science, 646 Swift Road West Point, NY, 10996-1905*
[2.]*University of Michigan, Electrical and Computer Engineering, 2600 Bonisteel Blvd. Ann Arbor, Michigan 48109, USA*
*\*z.zehnpfennig@us.army.mil*

**Abstract:** Stimulated Brillouin scattering recently allowed experimental excitation of surface acoustic resonances in micro-devices, enabling vibration at rates in the range of 50 MHz to 12 GHz. The experimental availability of such mechanical whispering gallery modes in photonic-MEMS raises questions on their structure and spectral distribution. Here we calculate the form and frequency of such vibrational surface whispering gallery modes, revealing diverse types of surface vibrations including longitudinal, transverse, and Rayleigh-type deformations. We parametrically investigate these various modes by changing their orders in the azimuthal, radial, and polar directions to reveal different vibrational structures including mechanical resonances that are localized near the interface with the environment where they can sense changes in the surroundings.

**1. Introduction**

Following an observation at St. Peter's Cathedral, Lord Rayleigh reported on circumferentially circulating acoustical resonances that he named "whispering gallery modes [1]." A century later, the optical counterpart of these acoustic whispering gallery modes is widely used in on-chip microdevices [2, 3] to allow Raman- [4] and Erbium- [5] lasers, parametric oscillations [6], third-harmonic generation [7], and excitation of mechanical breathing mode [8-10]. Recently, such optical whispering gallery modes were shown [11-13]to excite *mechanical whispering gallery modes* within microresonators, just like the ones that Rayleigh first observed within the cathedral. These mechanical vibrations were excited via the interplay between photoelastic scattering of light by sound and electrostrictive forces by light on sound. Interaction between sound and light is generally referred to as Brillouin scattering. When the excitation of the mechanical whispering gallery mode is from a single optical source, the mechanism is called Stimulated Brillouin Scattering.

Brillouin scattering has been studied with plane waves [8, 9, 14], vortices [15], and recently with mechanical whispering gallery modes [11-13]. In more detail: backward [8, 14] Brillouin scattering allowed excitation of mechanical whispering gallery modes in crystalline disks [12] as well as in amorphous silica micro-spheres [11] at rates above 10 GHz. Shortly afterward, forward [16-18] Brillouin scattering was suggested [19] and then used [13] in silica micro-resonators to optically excite whispering gallery vibrations at rates from 50 MHz to 1.4 GHz. The availability of the broad range of resonances from 50 MHz to above 10 GHz, their optical excitation and examination, as well as their whispering gallery propagation suggests a new degree of freedom in [10]photonic-MEMS with applications including sensitive detectors [19] and optomechanical oscillators operating at the quantum limit [20]. All of these studies are expected to benefit from low power consumption by these acoustic modes because their exciting Brillouin process is known to have the highest gain among all optical nonlinearities [14].

The acoustical density wave ($\tilde{\rho}$) and the optical waves ($\tilde{E}$) that are circulating in a sphere can be written in spherical coordinate system as

$$\tilde{\rho} = A_a(t)T_a(\theta,r)e^{i(M_{\phi a}\phi - \varpi_a t)},$$
$$\tilde{E}_p = A_p(t)T_p(\theta,r)e^{i(M_{\phi p}\phi - \varpi_p t)}, \quad (1)$$
$$\tilde{E}_S = A_S(t)T_S(\theta,r)e^{i(M_{\phi S}\phi - \varpi_S t)}.$$

Here $A$ stands for the wave amplitudes, $T$ for the wave distributions in the plane transverse with propagation, and $\varpi$ for the angular frequencies. Subscript $a$, $p$, and $S$ represent acoustical density, optical pump, and optical Stokes modes respectively. Similar to [15], $M_\phi$ corresponds to the angular momentum of the azimuthally circulating wave.

The electrostrictive excitation of the density mode by light is given by the acoustical wave equation with a source term driven by light on its right-hand side:

$$\frac{\partial^2 \tilde{\rho}}{\partial t^2} - V\nabla^2 \tilde{\rho} - b\frac{\partial}{\partial t}\nabla^2 \tilde{\rho} = \frac{1}{2}\varepsilon_0 \gamma_\theta \nabla^2 \left\langle \left(\tilde{E}_p + \tilde{E}_S\right)^2 \right\rangle. \quad (2)$$

Here $V$ is the speed of sound, $b$ represents dissipation, $\varepsilon_0$ is the vacuum permittivity, and $\gamma_\theta = \rho\frac{\partial s}{\partial \rho}$ is the electrostrictive coupling constant. The angle bracket on the right-hand side of Eq. (2) stands for time average over a period of several optical-cycle periods. A detailed modeling of such systems can be found in [21, 22]. In what follows we will numerically calculate the shape and frequency spectra of the mechanical whispering gallery modes (Eq. (1), $\tilde{\rho}$). Calculation of optical whispering gallery modes (Eq. (1), $\tilde{E}$) can be found in [23].

In broader opto-mechanical contexts, the mechanical whispering gallery modes [11-13] are excited by the compressive pressure of light. This is similar to other optical forces including centrifugal radiation pressure used to excite breathing modes [10, 24, 25], and gradient forces [26] that support excitation of flapping modes [27, 28]. The uniqueness of the whispering gallery modes is in offering a simple way to achieve a nearly material-limited mechanical quality factor. This is because the acoustical energy flows for the whispering gallery modes are orthogonal to the support so that no notch [29] or mechanical lambda-quarters [29, 30] are needed to prevent acoustical leakage. Additionally, the persistence of these modes near the interface with the environment makes them attractive to sense changes in their liquid or gaseous surroundings.

Generally speaking, the sensitivity of surface acoustic wave sensors [31] scales with their frequency squared. Though many challenges still exist, mechanical whispering gallery modes allow in this regard rates [11, 19] two orders of magnitude higher than in currently available surface acoustic wave sensors.

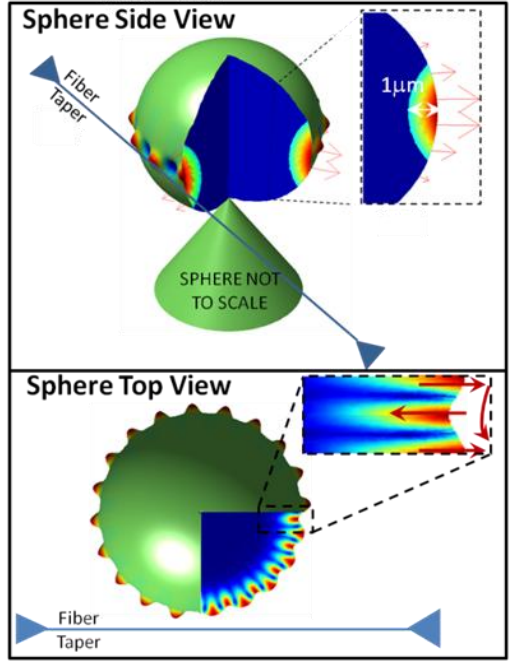

Fig. 1. Illustration of the mechanical whispering gallery resonance in a sphere. Deformation of the outer surface describes the exaggerated mechanical deformation. The cuts reveal also the internal deformation as indicated by colors.

Figure 1 illustrates the mechanical whispering gallery modes that we study here by describing a ~100-micron diameter sphere where the deformation is circumferentially circulating at the km/s-scale speed of sound. Each of the modes here will be indexed by $M_r, M_\theta$ and $M_\phi$ for the mode order in the corresponding directions using a spherical coordinate system. For example, the mode presented in Fig. 1 is $[M_r, M_\theta, M_\phi]$=[1,1,20], indicating 1st order in the radial and polar directions and 20 wavelengths are circumferentially resonating along the azimuthal direction. While the modes investigated here are traveling along $\phi$ to enhance acoustical power by multiple recirculation, they are standing-wave type in $\hat{r}$ and in the $\hat{\theta}$ directions. This is important since no propagation in $\hat{r}$ or $\hat{\theta}$ (contrary with [32, 33]) suggests here minimal leakage of energy through the support.

We note that another index can be added to describe either clockwise or counter-clockwise propagation of the mechanical whispering gallery mode.

**2. Calculated Rayleigh, transverse, and longitudinal whispering gallery modes**

Unlike optics, where modes are vectorial and usually restricted to choose between two polarization states, the mechanical strain is a tensor thereby allowing a variety of modes. For example, in earthquake research, waves of longitudinal-, transverse-, and Rayleigh-type are known. We calculate mechanical whispering gallery modes in a silica sphere similar to what were experimentally observed in [11, 13]. In our first example, the circumference is equal to 20 in units of acoustical wavelength. Similar to earthquake waves, the calculated modes include deformations of Rayleigh [34], transverse, and longitudinal type, as shown in Fig. 2. The deformation for these modes is along the azimuthal, polar, and radial-polar directions, as indicated by arrows. The Rayleigh mode is unique in that a particle follows an elliptical path. This can be useful in applying forces to move surrounding fluids [35] and might create microvortices around our structure. We make the calculations with a finite-element software [36]. For the Rayleigh and transverse modes there were no constraints on the structure except for the symetries of the problem. However, in calculating the longitudinal mode, it was difficult to find the solution since it was surrounded by many high-order Rayleigh and transverse modes. This is because the speed of the longitudinal wave is much faster than the other waves. We therefore forced longitudinal motion on several structure elements for calculating the longitudinal mode. We know that this approximation is not perfect as the Poisson ratio suggests that even a longitudinal mode has a residual transverse deformation. This aproximation is justified by experimental results [11] very close to this theoretical calculation. Another indication for the validity of our aproximation is that our solution asymptotically converges to the analytical (longitudinal) speed of sound in bulk when the radius of the sphere is much larger than wavelength.

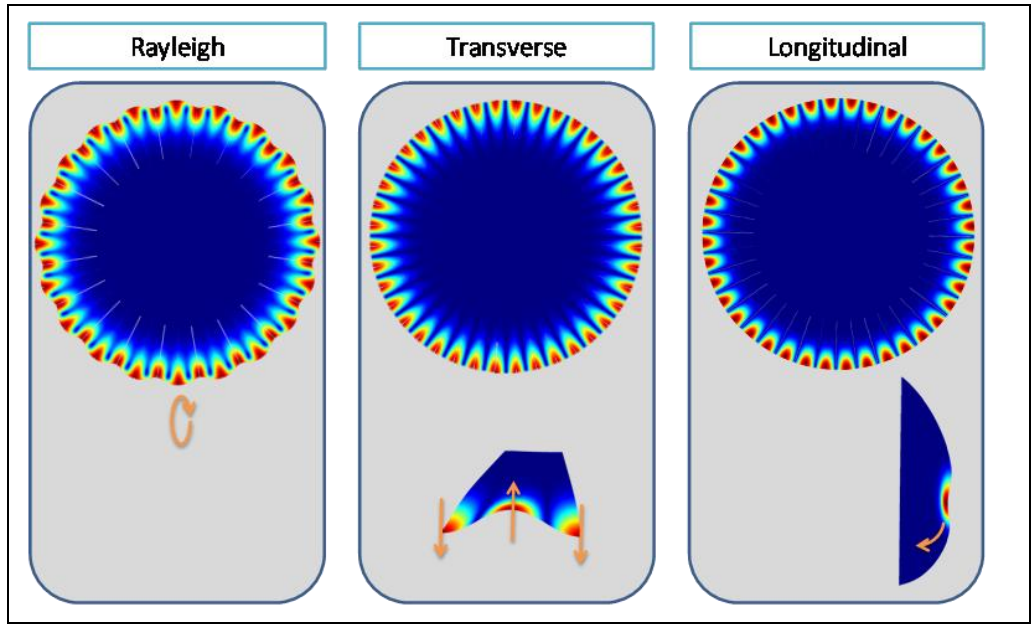

Fig. 2. Rayleigh, transverse, and longitudinal whispering gallery modes where energy is confined within a wavelength distance from the interface. The major deformation is either radial, polar, or azimuthal as indicated by arrows. $M_\phi$ in this calculation is 20, indicating that 20 acoustical wavelengthes are resonating along the spehre circumference. Color represents absolute value of the deformation.

Looking at Fig. 2, we see that the curvature of the sphere confines all of the mechanical whispering gallery modes to within a wavelength distance of the interface; this was verified for modes having between 10 and 2000 wavelengths along the circumference. We therefore define all of the resonances studied here as surface-acoustic type. All waves here are being confined to the interface. This is different from acoustic propagation on planar interfaces where only the Rayleigh wave is of a surface type.

**3. High-order whispering gallery modes**

Each of the modes calculated above (Rayleigh, transverse, and longitudinal) represents a family of modes. Each family has many members with different order in all three directions. In what follows we will perform a parametric study whereby we observe the effect of varying azimuthal mode number, $M_\phi$, and the transverse mode numbers ($M_q$, $M_\rho$) on the structure and frequency of the whispering gallery mode.

### 3.1 Increasing mode order in a direction transverse to propagation

We will start by increasing the mode order in the $\hat{r}$ and $\hat{\theta}$ directions. These directions are transverse to propagation. Figure 3 shows that, much like their optical counterpart [37, 38], a mechanical whispering gallery mode can be high-order in the polar and radial directions.

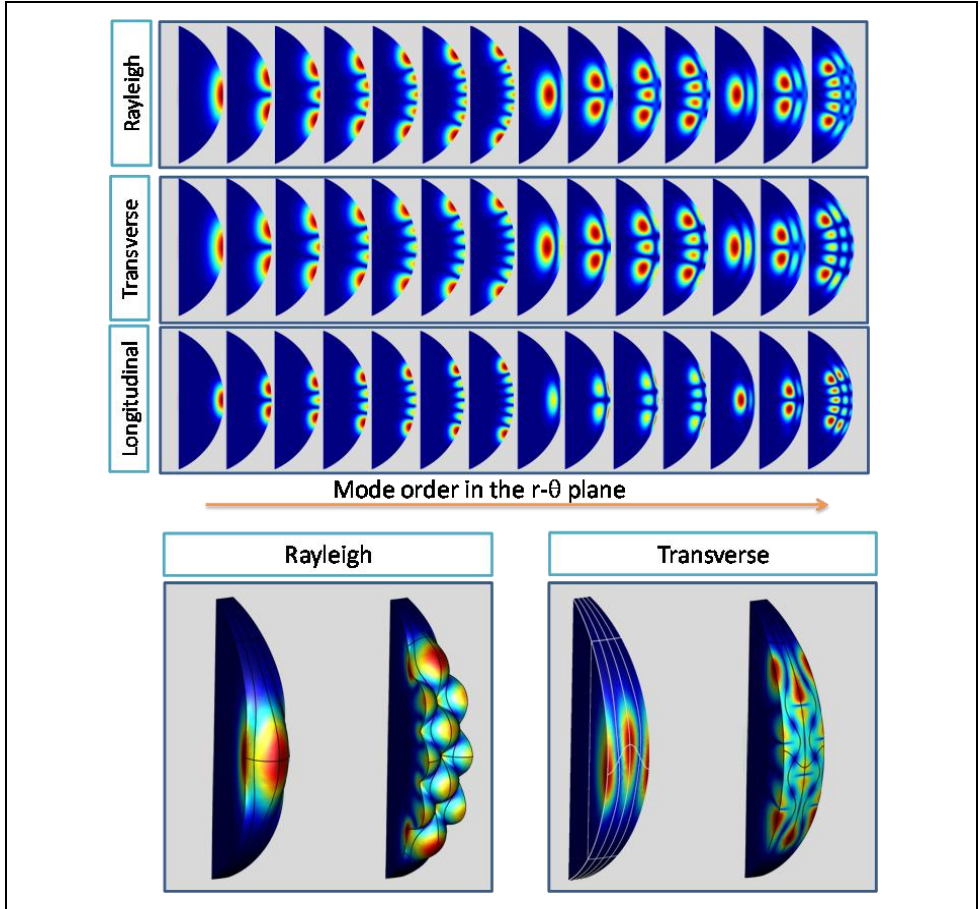

Fig. 3: High-order mechanical whispering gallery modes. Top: Increasing the mode order in the radial and polar directions for mechanical whispering gallery modes in a silica sphere. Color represents deformation. $M_\phi$=20. Bottom: We depict several of the top modes and present them in 3D. The presented section is one acoustic wavelength in the azimuthal direction. The equator is seen to deform into a sine where the Rayleigh mode the deformation is in the radial direction and for the transverse mode the deformation is in the polar direction.

In Fig. 3 we can see that mode $[M_r, M_\theta, M_\phi]$=[3,1,20] has minimal deformation on the interface and is hence attractive for applications where minimal dissipation by air is required. On the contrary, the $[M_r, M_\theta, M_\phi]$=[1,7,20] is extending into a large area on the interface; this fact might be useful for sensing changes in the surroundings.

### 3.2 Increasing Mode Order in Directions Parallel to Propagation.

The azimuthal order of the experimentally excited modes for a typical 100-micron radius sphere varies from $M_\phi$=600 for backward Brillouin scattering excitation [11] and down to $M_\phi$~10 [13] when the excitation is by forward scattering. In what follows, we accordingly change $M_\phi=$ along this span of azimuthal wavelengths.

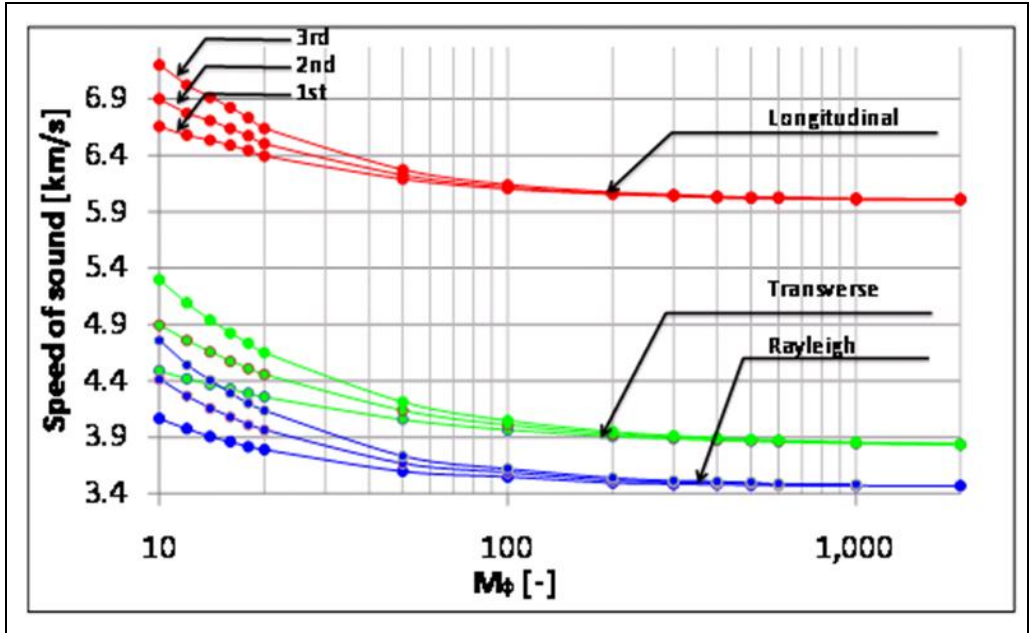

Fig. 4. High-order mechanical whispering gallery modes. The calculated speed of sound is shown as a function of the azimuthal mode order, $M_\phi$. The first three transverse orders are given for the Rayleigh-, transverse-, and longitudinal families. At large $M_\phi$, the speed of sound asymptotically converges to the relevant speed of sound in bulk media (see Table 1).

Figure 4 describes the speed of sound (at interface) for $M_\phi$=10 to 2000. For a sphere with radius, $r$, and azimuthal mode order, $M_\phi$, the resonance frequency can be easily calculated by dividing the relevant speed of sound by one acoustical wavelength: $2\pi r / M_\theta$. As expected, for $M_\phi$ above 200, the speeds of sound asymptotically converge to the bulk speeds of sound. This is shown in Table 1 where we compare between the analytical [39-41] and numerical value for the speeds of sound. The analysis suggests that for cases where $M_\phi > 200$, the bulk speed of sound can be used as a good approximation for calculating the resonant frequency.

| Bulk SiO$_2$ | | Numerical calculation, SiO$_2$ Sphere, M phi=2000 | | |
|---|---|---|---|---|
| Wave | Velocity [m/s] | Mode | Deformation | Velocity [m/s] |
| Longitudinal | $V_L = \left(\dfrac{E(v-1)}{\rho(2v^2+v-1)}\right)^{1/2} = 5972$ | Longitudinal | Azimuthal | 5957 |
| Transverse | $V_T = \left(\dfrac{E}{2\rho(v+1)}\right)^{1/2} = 3766$ | Transverse | Polar | 3787 |
| Rayleigh | $V_R = \dfrac{V_T(0.87+1.12v)}{(1+v)} = 3413$ | Rayleigh | Radial-Polar | 3420 |

Table 1: Analytically and numerically calculated speeds of sound. For a large sphere, the speeds of sound converge to bulk speeds. Error between the analytically and numerically calculated speeds is less than 1%. Here $\rho$ is the medium density at mechanical equilibrium, $E$ it the Young modulus and $\nu$ is the Poisson modulus. Equations taken from [39-41].

In order to give a scale for the resonant frequencies and their dependency on $M_\phi$, we will now present the results from Fig. 4 in terms of frequencies for a $r = 100$-micron silica sphere. Figure 5 shows these vibrational frequencies. Regions where resonances were experimentally optically excited in silica spheres [11, 13] are marked by shadows. The shadows extend to higher frequencies representing the effects of the high order transverse modes in accordance with observation.

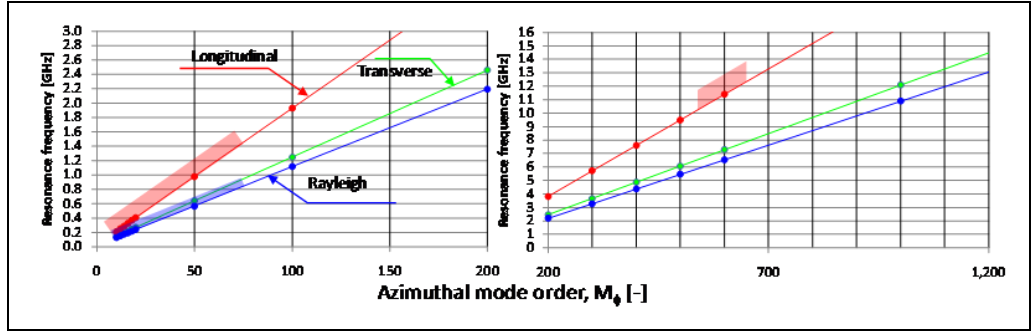

Fig. 5. Vibration frequencies for the various modes in a $r=100$ micron silica sphere as a function of their azimuthal mode order. Left, with $M_\phi$ typical to forward Brillouin excitation. Right, with $M_\phi$ typical to backward Brillouin excitation. The shadowed regions estimate how high resonance frequencies can go for each of these modes via relying on high order transverse members of this mode family. The shadowed region is bounded in the $M_\phi$ direction as estimation from momentum conservation consideration. We assume excitation with 1.5-micron telecom pump.

**4. Conclusion**

Mechanical whispering gallery modes in a silica microsphere were calculated here to reveal a variety of deformations that are now experimentally possible in photonic-MEMS. As expected – as with previous studies in bulk materials - deformations including longitudinal-, transverse-, and Rayleigh-type resonances were calculated in the microsphere. However, while in bulk materials only the Rayleigh type deformation is a surface wave, here all of the deformation types are confined near the surface. Further, each of these modal families has members of different azimuthal, radial, and polar orders. Modes with different extension into the interface were calculated to support either enhancement or reduction of environmental effects upon need. We believe the new capacity to optically excite mechanical whispering gallery modes in microdevices will benefit from the numerical calculation presented here of the frequency, shape, and structure of these vibrations. For example, the strain distribution of these modes allow calculating their modal overlap with the optical modes [23], calculating their photoelastic index modulation [42, 43], and calculating the amplitude of their Brownian fluctuation by comparison of their strain energy with the Boltzmann constant multiplied by temperature.

**References and links**